\documentclass[aps,pra,twocolumn,showpacs,floatfix,superscriptaddress]{revtex4-1}

  \usepackage[utf8]{inputenc}
  \usepackage[T1]{fontenc}
  \usepackage{float} 
  \usepackage{color}  

\usepackage{graphicx}
\usepackage{amsmath}
\usepackage{amssymb}
\usepackage{bbm}
\usepackage{indentfirst}
\usepackage{dcolumn}
\usepackage{soul}
\usepackage{mathrsfs}
\usepackage{bigints}
\usepackage{relsize}
\usepackage{tikz}
\usetikzlibrary{calc}
\usetikzlibrary{arrows.meta}

\begin{document}

\title{Localized spin waves in isolated $k\pi$ skyrmions}

\author{Levente R\'{o}zsa}
\email{rozsa.levente@physnet.uni-hamburg.de}
\author{Julian Hagemeister}
\author{Elena Y. Vedmedenko}
\author{Roland Wiesendanger}
\affiliation{Department of Physics, University of Hamburg, D-20355 Hamburg, Germany}
\date{\today}
\pacs{}

\begin{abstract}

The localized magnon modes of isolated $k\pi$ skyrmions on a field-polarized background are analyzed based on the Landau--Lifshitz--Gilbert equation within the terms of an atomistic classical spin model, with system parameters based on the Pd/Fe biatomic layer on Ir(111). For increasing skyrmion order $k$ a higher number of excitation modes are found, including modes with nodes in the radial eigenfunctions. It is shown that at low fields $2\pi$ and $3\pi$ skyrmions are destroyed via a burst instability connected to a breathing mode, while $1\pi$ skyrmions undergo an elliptic instability. At high fields all $k\pi$ skyrmions collapse due to the instability of a breathing mode. The effective damping parameters of the spin waves are calculated in the low Gilbert damping limit, and they are found to diverge in the case of the lowest-lying modes at the burst and collapse instabilities, but not at the elliptic instability. It is shown that the breathing modes of $k\pi$ skyrmions may become overdamped at higher Gilbert damping values.

\end{abstract}

\maketitle

\section{Introduction}

Magnetic skyrmions are localized particle-like spin configurations \cite{Bogdanov}, which have become the focus of intense research activities over the last years due to their promising applications in spintronic devices \cite{Fert,Jiang,Hsu,Buttner}. While their particle-like properties make them suitable to be used as bits of information, the collective excitations of the spins constituting the magnetic skyrmion, known as spin waves or magnons, open possible applications in the field of magnonics \cite{Garst}.

These spin wave modes were first investigated theoretically \cite{Petrova,Zang,Mochizuki,dosSantos} and experimentally \cite{Schwarze,Onose,Ehlers} in skyrmion lattice phases, where the interactions between the skyrmions lead to the formation of magnon bands. If a skyrmion is confined in a finite-sized nanoelement, it will possess discrete excitation frequencies \cite{Gareeva,Kim,Beg,Mruczkiewicz}. Although such geometries have also been successfully applied to the time-resolved imaging of the dynamical motion of magnetic bubble domains \cite{Buttner2,Finizio}, in such a case it is not possible to distinguish between the excitations of the particle-like object itself and spin waves forming at the edges of the sample \cite{Gareeva}. In order to rule out boundary effects, the excitations of isolated skyrmions have to be investigated, as was performed theoretically in Refs.~\cite{Lin,Schutte,Kravchuk,Rozsa}. It was suggested recently \cite{Satywali} that the experimentally determined excitation frequencies in the Ir/Fe/Co/Pt multilayer system may be identified as spin wave modes of isolated skyrmions, rather than as magnons stemming from an ordered skyrmion lattice.

In most investigations skyrmions correspond to simple domains with the magnetization in their core pointing opposite to the collinear background. However, it was shown already in Ref.~\cite{Bogdanov4} that the Dzyaloshinsky--Moriya interaction \cite{Dzyaloshinsky,Moriya} responsible for their stabilization may also lead to the formation of structures where the direction of the magnetization rotates multiple times between the center of the structure and the collinear region. Such target states or $k\pi$ skyrmions, where $k$ is the number of sign changes of the out-of-plane magnetization when moving along the radial direction, have also been investigated in constricted geometries \cite{Rohart,Leonov,Mulkers,Beg2,Liu}. The experimental observation of localized spin structures with multiple rotations has been mainly restricted to systems with negligible Dzyaloshinsky--Moriya interaction so far \cite{Finazzi,Streubel,Finizio}, where the formation of domain structures is attributed to the magnetostatic dipolar interaction.

The collapse of isolated $k\pi$ skyrmions and their creation in nanodots by switching the external field direction was recently investigated in Ref.~\cite{Hagemeister}. It was found that during the creation process the skyrmions display significant size oscillations resembling breathing eigenmodes. In Ref.~\cite{Bogdanov4}, the stability of $k\pi$ skyrmions was studied in a system with a ferromagnetic ground state, and it was found that applying the external field opposite to the background magnetization leads to a divergence of the skyrmion radius at a critical field value, a so-called burst instability. This instability can be attributed to a sign change of one of the eigenvalues of the energy functional expanded around the $k\pi$ skyrmion configuration, intrinsically related to the dynamics of the system. However, the spin wave frequencies of isolated $k\pi$ skyrmions remain unexplored.

Besides the excitation frequencies themselves, the lifetime of spin waves is also of crucial importance in magnonics applications. This is primarily influenced by the Gilbert damping parameter $\alpha$ \cite{Gilbert}, the value of which can be determined experimentally based on resonance lineshapes measured in the collinear state \cite{Schwarze,Satywali,Finizio}. It was demonstrated recently \cite{Rozsa} that the noncollinear spin structure drastically influences the effective damping parameter acting on the spin waves, leading to mode-dependent and enhanced values compared to the Gilbert damping parameter. This effect was discussed through the example of the $1\pi$ skyrmion in Ref.~\cite{Rozsa}, but it is also expected to be observable for $k\pi$ skyrmions with higher order $k$.

Here the localized spin wave frequencies of isolated $k\pi$ skyrmions are investigated in a classical atomistic spin model. The parameters in the Hamiltonian represent the Pd/Fe/Ir(111) model-type system, where the properties of skyrmions have been studied in detail both from the experimental \cite{Romming2,Romming} and from the theoretical \cite{Dupe,Simon,Hagemeister2,Hagemeister} side. The paper is organized as follows. The classical atomistic spin Hamiltonian and the method of calculating the eigenmodes is introduced in Sec.~\ref{sec2a}, while the angular momentum and nodal quantum numbers characterizing the excitations are defined in Sec.~\ref{sec2b} within the framework of the corresponding micromagnetic model. Eigenfrequencies equal to or approaching zero are discussed in Sec.~\ref{sec2c}, and the effective damping parameters are introduced in Sec.~\ref{sec2e}. The eigenmodes of $k\pi$ skyrmions with $k=1,2,3$ are compared in Sec.~\ref{sec3a}, the instabilities occurring at low and high field values are discussed in connection to magnons with vanishing frequencies in Sec.~\ref{sec3b}, and the effective damping parameters of the different modes are calculated for vanishing and higher values of the Gilbert damping in Secs.~\ref{sec3c} and \ref{sec3d}, respectively. A summary is given in Sec.~\ref{sec4}.

\section{Methods}

\subsection{Atomistic model\label{sec2a}}

The system is described by the classical atomistic model Hamiltonian
\begin{eqnarray}
H&=&-\frac{1}{2}\sum_{\left<i,j\right>}
J\boldsymbol{S}_{i}\boldsymbol{S}_{j}-\frac{1}{2}\sum_{\left<i,j\right>}\boldsymbol{D}_{ij}\left(\boldsymbol{S}_{i}\times\boldsymbol{S}_{j}\right)\nonumber
\\
&&-\sum_{i}K\left(S_{i}^{z}\right)^{2}-\sum_{i}\mu_{\textrm{s}}\boldsymbol{B}\boldsymbol{S}_{i},\label{eqn1}
\end{eqnarray}
with the $\boldsymbol{S}_{i}$ unit vectors representing the spins in a single-layer triangular lattice; $J$, $\boldsymbol{D}_{ij}$, and $K$ denoting nearest-neighbor Heisenberg and Dzyaloshinsky--Moriya exchange interactions and on-site magnetocrystalline anisotropy, respectively; while $\mu_{\textrm{s}}$ and $\boldsymbol{B}$ stand for the spin magnetic moment and the external magnetic field. The numerical values of the parameters are taken from Ref.~\cite{Hagemeister}, being $J=5.72\,\textrm{meV}, D=\left|\boldsymbol{D}_{ij}\right|=1.52\,\textrm{meV}, K=0.4\,\textrm{meV}$, and $\mu_{\textrm{s}}=3\mu_{\textrm{B}}$, describing the Pd/Fe/Ir(111) system. The energy parameters were determined based on measuring the field-dependence of $1\pi$ skyrmion profiles in the system by spin-polarized scanning tunneling microscopy in Ref.~\cite{Romming}.

During the calculations the external field $\boldsymbol{B}$ is oriented along the out-of-plane $z$ direction. The equilibrium $k\pi$ skyrmion structures are determined from a reasonable initial configuration by iteratively rotating the spins $\boldsymbol{S}_{i}$ towards the direction of the effective magnetic field $\boldsymbol{B}_{i}^{\textrm{eff}}=-\dfrac{1}{\mu_{\textrm{s}}}\dfrac{\partial H}{\partial \boldsymbol{S}_{i}}$. The iteration is performed until the torque acting on the spins, $\boldsymbol{T}_{i}=-\boldsymbol{S}_{i}\times\boldsymbol{B}_{i}^{\textrm{eff}}$, becomes smaller at every lattice site than a predefined value, generally chosen to be $10^{-8}\,\textrm{meV}/\mu_{\textrm{B}}$. The calculations are performed on a lattice with periodic boundary conditions, with system sizes up to $256 \times 256$ for the largest $k\pi$ skyrmions in order to avoid edge effects and enable the accurate modeling of isolated skyrmions.

Once the equilibrium configuration $\boldsymbol{S}_{i}^{(0)}$ is determined, the spins are rotated to a local coordinate system $\tilde{\boldsymbol{S}}_{i}=\boldsymbol{R}_{i}\boldsymbol{S}_{i}$ using the rotational matrices $\boldsymbol{R}_{i}$. In the local coordinate system the equilibrium spin directions are pointing along the local $z$ axis, $\tilde{\boldsymbol{S}}_{i}^{(0)}=\left(0,0,1\right)$. The Hamiltonian in Eq.~(\ref{eqn1}) is expanded up to second-order terms in the small variables $\tilde{S}_{i}^{x},\tilde{S}_{i}^{y}$ as (cf. Ref.~\cite{Rozsa})
\begin{eqnarray}
H&\approx&H_{0}+\frac{1}{2}\left(\tilde{\boldsymbol{S}}^{\perp}\right)^{T}\boldsymbol{H}_{\textrm{SW}}\tilde{\boldsymbol{S}}^{\perp}\nonumber
\\
&=&H_{0}+\frac{1}{2}\left[\begin{array}{cc} \tilde{\boldsymbol{S}}^{x} & \tilde{\boldsymbol{S}}^{y} \end{array}\right]\left[\begin{array}{cc} \boldsymbol{A}_{1} & \boldsymbol{A}_{2} \\ \boldsymbol{A}_{2}^{\dag} & \boldsymbol{A}_{3} \end{array}\right]\left[\begin{array}{c} \tilde{\boldsymbol{S}}^{x} \\ \tilde{\boldsymbol{S}}^{y} \end{array}\right].\label{eqn2}
\end{eqnarray}

The matrix products are understood to run over lattice site indices $i$, with the matrix components reading
\begin{eqnarray}
A_{1,ij}&=&-\tilde{J}_{ij}^{xx}+\delta_{ij}\left(\sum_{k}\tilde{J}_{ik}^{zz}-2\tilde{K}_{i}^{xx}+2\tilde{K}_{i}^{zz}+\mu_{\textrm{s}}\tilde{B}_{i}^{z}\right),\:\:\:\:\:\:\label{eqn3}
\\
A_{2,ij}&=&-\tilde{J}_{ij}^{xy}-\delta_{ij}2\tilde{K}_{i}^{xy},\label{eqn4}
\\
A_{3,ij}&=&-\tilde{J}_{ij}^{yy}+\delta_{ij}\left(\sum_{k}\tilde{J}_{ik}^{zz}-2\tilde{K}_{i}^{yy}+2\tilde{K}_{i}^{zz}+\mu_{\textrm{s}}\tilde{B}_{i}^{z}\right).\:\:\:\:\:\:\label{eqn5}
\end{eqnarray}

The energy terms in the Hamiltonian are rotated to the local coordinate system via $\tilde{\boldsymbol{J}}_{ij}=\boldsymbol{R}_{i}\left[J\boldsymbol{I}-\boldsymbol{D}_{ij}\times\right]\boldsymbol{R}_{j}^{T}, \tilde{\boldsymbol{K}}_{i}=\boldsymbol{R}_{i}\boldsymbol{K}\boldsymbol{R}_{j}^{T},$ and $\tilde{\boldsymbol{B}}_{i}=\boldsymbol{R}_{i}\boldsymbol{B}$, where $\boldsymbol{I}$ is the $3\times 3$ identity matrix, $\boldsymbol{D}_{ij}\times$ is the matrix describing the vector product with $\boldsymbol{D}_{ij}$, and $\boldsymbol{K}$ is the anisotropy matrix with the only nonzero element being $K^{zz}=K$. 

The spin wave frequencies are obtained from the linearized Landau--Lifshitz--Gilbert equation \cite{Landau,Gilbert}
\begin{eqnarray}
\partial_{t}\tilde{\boldsymbol{S}}^{\perp}=\frac{\gamma'}{\mu_{\textrm{s}}}\left(-\textrm{i}\boldsymbol{\sigma}^{y}-\alpha\right)\boldsymbol{H}_{\textrm{SW}}\tilde{\boldsymbol{S}}^{\perp}=\boldsymbol{D}_{\textrm{SW}}\tilde{\boldsymbol{S}}^{\perp},\label{eqn6}
\end{eqnarray}
with $\boldsymbol{\sigma}^{y}=\left[\begin{array}{cc} 0 & -\textrm{i}\boldsymbol{I}_{\textrm{s}} \\ \textrm{i}\boldsymbol{I}_{\textrm{s}} & 0 \end{array}\right]$ the Pauli matrix in Cartesian components and acting as the identity matrix $\boldsymbol{I}_{\textrm{s}}$ in the lattice site summations. The symbol $\gamma'$ denotes the gyromagnetic ratio $\gamma=\frac{ge}{2m}$ divided by a factor of $1+\alpha^{2}$, with $g$ the electron $g$ factor, $e$ the elementary charge, $m$ the electron's mass, and $\alpha$ the Gilbert damping parameter. Equation~(\ref{eqn6}) is rewritten as an eigenvalue equation by assuming the time dependence $\tilde{\boldsymbol{S}}^{\perp}\left(t\right)=\textrm{e}^{-\textrm{i}\omega_{q}t}\tilde{\boldsymbol{S}}_{q}^{\perp}$ and performing the replacement $\partial_{t}\rightarrow -\textrm{i}\omega_{q}$.

Since the $k\pi$ skyrmions represent local energy minima, $\boldsymbol{H}_{\textrm{SW}}$  in Eq.~(\ref{eqn2}) is a positive semidefinite matrix. 
For $\alpha=0$ the $\omega_{q}$ frequencies of $\boldsymbol{D}_{\textrm{SW}}$ are real and they always occur in $\pm\omega_{q}$ pairs on the subspace where $\boldsymbol{H}_{\textrm{SW}}$ is strictly positive, for details see, e.g., Ref.~\cite{Rozsa}. In the following, we will only treat the solutions with $\textrm{Re}\:\omega_{q}>0$, but their $\textrm{Re}\:\omega_{q}<0$ pairs are also necessary for constructing real-valued eigenvectors of Eq.~(\ref{eqn6}). The zero eigenvalues are discussed in Sec.~\ref{sec2c}.

As is known from previous calculations for $1\pi$ skyrmions \cite{Schutte,Kravchuk,Rozsa}, the localized excitation modes of $k\pi$ skyrmions are found below the ferromagnetic resonance frequency $\omega_{\textrm{FMR}}=\frac{\gamma}{\mu_{\textrm{s}}}\left(2K+\mu_{\textrm{s}}B\right)$. During the numerical solution of Eq.~(\ref{eqn6}) these lowest-lying eigenmodes of the sparse matrix $\boldsymbol{D}_{\textrm{SW}}$ are determined, as implemented in the \textsc{montecrystal} atomistic spin simulation program \cite{MonteCrystal}.


\subsection{Micromagnetic model\label{sec2b}}

The atomistic model described in the previous Section enables the treatment of noncollinear spin structures where the direction of the spins significantly differs between neighboring lattice sites. This is especially important when discussing the collapse of $k\pi$ skyrmions on the lattice as was performed in Ref.~\cite{Hagemeister}. Here we will discuss the micromagnetic model which on the one hand is applicable only if the characteristic length scale of noncollinear structures is significantly larger than the lattice constant, but on the other hand enables a simple classification of the spin wave modes.

The free energy functional of the micromagnetic model is defined as
\begin{align}
\mathcal{H}=&\int\mathcal{A}\sum_{\alpha=x,y,z}\left(\boldsymbol{\nabla}S^{\alpha}\right)^{2}+\mathcal{K}\left(S^{z}\right)^{2}-\mathcal{M}BS^{z}\nonumber
\\
&+\mathcal{D}\left(S^{z}\partial_{x}S^{x}-S^{x}\partial_{x}S^{z}+S^{z}\partial_{y}S^{y}-S^{y}\partial_{y}S^{z}\right)\textrm{d}\boldsymbol{r},\:\:\:\:\:\:\label{eqn9}
\end{align}
where for the Pd/Fe/Ir(111) system the following parameter values were used: $\mathcal{A}=2.0\,\textrm{pJ/m}$ is the exchange stiffness, $\mathcal{D}=-3.9\,\textrm{mJ/m}^{2}$ is the Dzyaloshinsky--Moriya interaction describing right-handed rotation \cite{Dupe}, $\mathcal{K}=-2.5\,\textrm{MJ/m}^{3}$ is the easy-axis anisotropy, and $\mathcal{M}=1.1\,\textrm{MA/m}$ is the saturation magnetization.

The equilibrium spin structure $\boldsymbol{S}^{(0)}=\left(\sin\Theta_{0}\cos\Phi_{0},\sin\Theta_{0}\sin\Phi_{0},\cos\Theta_{0}\right)$ of $k\pi$ skyrmions will be cylindrically symmetric, given by $\Phi_{0}\left(r,\varphi\right)=\varphi+\pi$ due to the right-handed rotational sense and $\Theta_{0}\left(r,\varphi\right)=\Theta_{0}\left(r\right)$, which is the solution of the Euler--Lagrange equation
\begin{eqnarray}
&&\mathcal{A}\left(\partial_{r}^{2}\Theta_{0}+\frac{1}{r}\partial_{r}\Theta_{0}-\frac{1}{r^{2}}\sin\Theta_{0}\cos\Theta_{0}\right)+\left|\mathcal{D}\right|\frac{1}{r}\sin^{2}\Theta_{0}\nonumber
\\
&&+\mathcal{K}\sin\Theta_{0}\cos\Theta_{0}-\frac{1}{2}\mathcal{M}B\sin\Theta_{0}=0.\label{eqn10}
\end{eqnarray}

The skyrmion order $k$ is encapsulated in the boundary conditions $\Theta_{0}\left(0\right)=k\pi,\Theta_{0}\left(\infty\right)=0$. Equation~(\ref{eqn10}) is solved numerically in a finite interval $r\in[0,R]$ significantly larger than the equilibrium $k\pi$ skyrmion size. A first approximation to the spin structure is constructed based on the corresponding initial value problem using the shooting method \cite{Bogdanov4}, then iteratively optimizing the structure using a finite-difference discretization.

The spin wave Hamiltonian may be determined analogously to Eq.~(\ref{eqn2}), by using the local coordinate system $\Theta=\Theta_{0}+\tilde{S}^{x}, \Phi=\Phi_{0}+\frac{1}{\sin\Theta_{0}}\tilde{S}^{y}$. The matrices in Eqs.~(\ref{eqn3})-(\ref{eqn5}) are replaced by the operators
\begin{eqnarray}
A_{1}&=&-2\mathcal{A}\left(\boldsymbol{\nabla}^{2}-\frac{1}{r^{2}}\cos2\Theta_{0}\right)-2\left|\mathcal{D}\right|\frac{1}{r}\sin2\Theta_{0}\nonumber
\\
&&-2\mathcal{K}\cos2\Theta_{0}+\mathcal{M}B\cos\Theta_{0},\label{eqn11}
\\
A_{2}&=&4\mathcal{A}\frac{1}{r^{2}}\cos\Theta_{0}\partial_{\varphi}-2\left|\mathcal{D}\right|\frac{1}{r}\sin\Theta_{0}\partial_{\varphi},\label{eqn12}
\\
A_{3}&=&-2\mathcal{A}\left\{\boldsymbol{\nabla}^{2}+\left[\left(\partial_{r}\Theta_{0}\right)^{2}-\frac{1}{r^{2}}\cos^{2}\Theta_{0}\right]\right\}\nonumber
\\
&&-2\left|\mathcal{D}\right|\left(\partial_{r}\Theta_{0}+\frac{1}{r}\sin\Theta_{0}\cos\Theta_{0}\right)\nonumber
\\
&&-2\mathcal{K}\cos^{2}\Theta_{0}+\mathcal{M}B\cos\Theta_{0}.\label{eqn13}
\end{eqnarray}

Due to the cylindrical symmetry of the structure, the solutions of Eq.~(\ref{eqn6}) are sought in the form $\tilde{\boldsymbol{S}}^{\perp}\left(r,\varphi,t\right)=\textrm{e}^{-\textrm{i}\omega_{n,m}t}\textrm{e}^{\textrm{i}m\varphi}\tilde{\boldsymbol{S}}_{n,m}^{\perp}\left(r\right)$, performing the replacements $\partial_{t}\rightarrow -\textrm{i}\omega_{n,m}$ and $\partial_{\varphi}\rightarrow \textrm{i}m$. For each angular momentum quantum number $m$, an infinite number of solutions indexed by $n$ may be found, but only a few of these are located below $\omega_{\textrm{FMR}}=\frac{\gamma}{\mathcal{M}}\left(-2\mathcal{K}+\mathcal{M}B\right)$, hence representing localized spin wave modes of the $k\pi$ skyrmions. The different $n$ quantum numbers typically denote solutions with different numbers of nodes, analogously to the quantum-mechanical eigenstates of a particle in a box.

Because of the property $\boldsymbol{H}_{\textrm{SW}}\left(m\right)=\boldsymbol{H}^{*}_{\textrm{SW}}\left(-m\right)$ and $\boldsymbol{H}_{\textrm{SW}}$ being self-adjoint, the eigenvalues of $\boldsymbol{H}_{\textrm{SW}}\left(m\right)$ and $\boldsymbol{H}_{\textrm{SW}}\left(-m\right)$ coincide, leading to a double degeneracy apart from the $m=0$ modes. The $\pm\omega_{q}$ eigenvalue pairs of $\boldsymbol{D}_{\textrm{SW}}$ discussed in Sec.~\ref{sec2a} for the atomistic model at $\alpha=0$ in this case can be written as $\omega_{n,m}=-\omega_{n,-m}$. However, considering only the modes with $\textrm{Re}\:\omega_{n,m}>0$, one has $\omega_{n,m} \neq \omega_{n,-m}$ indicating nonreciprocity or an energy difference between clockwise ($m<0$) and counterclockwise ($m>0$) rotating modes \cite{Mruczkiewicz,Rozsa}.

%

For finding the eigenvectors and eigenvalues of the micromagnetic model, Eq.~(\ref{eqn6}) is solved using a finite-difference method on the $r\in[0,R]$ interval. For treating the Laplacian $\boldsymbol{\nabla}^{2}$ in Eqs.~(\ref{eqn11}) and (\ref{eqn13}) the improved discretization scheme suggested in Ref.~\cite{Laliena} was applied, which enables a more accurate treatment of modes with eigenvalues converging to zero in the infinite and continuous micromagnetic limit.

The spin wave modes of the atomistic model discussed in Sec.~\ref{sec2a} were assigned the $\left(n,m\right)$ quantum numbers, which are strictly speaking only applicable in the micromagnetic limit with perfect cylindrical symmetry, by visualizing the real-space structure of the numerically obtained eigenvectors.

\subsection{Goldstone modes and instabilities\label{sec2c}}

Since the translation of the $k\pi$ skyrmions on the collinear background in the plane costs no energy, the spin wave Hamiltonian $\boldsymbol{H}_{\textrm{SW}}$ possesses two eigenvectors belonging to zero eigenvalue, representing the Goldstone modes of the system. Within the micromagnetic description of Sec.~\ref{sec2b}, these may be expressed analytically as \cite{Schutte,Kravchuk,Rozsa}
\begin{eqnarray}
\left(\tilde{S}^{x},\tilde{S}^{y}\right)=\textrm{e}^{-\textrm{i}\varphi}\left(-\partial_{r}\Theta_{0},\textrm{i}\frac{1}{r}\sin\Theta_{0}\right),\label{eqn14}
\\
\left(\tilde{S}^{x},\tilde{S}^{y}\right)=\textrm{e}^{\textrm{i}\varphi}\left(-\partial_{r}\Theta_{0},-\textrm{i}\frac{1}{r}\sin\Theta_{0}\right).\label{eqn15}
\end{eqnarray}

Equations (\ref{eqn14}) and (\ref{eqn15}) represent eigenvectors of the dynamical matrix $\boldsymbol{D}_{\textrm{SW}}$ as well. From Eqs.~(\ref{eqn2}) and (\ref{eqn6}) it follows that the eigenvectors of $\boldsymbol{H}_{\textrm{SW}}$ and $\boldsymbol{D}_{\textrm{SW}}$ belonging to zero eigenvalue must coincide, $\boldsymbol{H}_{\textrm{SW}}\tilde{\boldsymbol{S}}^{\perp}=\boldsymbol{0} \Leftrightarrow \boldsymbol{D}_{\textrm{SW}}\tilde{\boldsymbol{S}}^{\perp}=\boldsymbol{0}$, because $\left(-\textrm{i}\boldsymbol{\sigma}^{y}-\alpha\right)$ in Eq.~(\ref{eqn6}) is an invertible matrix. Because from the solutions of the equation of motion (\ref{eqn6}) we will only keep the ones satisfying $\textrm{Re}\:\omega_{n,m}>0$, the eigenvectors from Eqs. (\ref{eqn14}) and (\ref{eqn15}) will be denoted as the single spin wave mode $\omega_{0,-1}=0$.

Since the eigenvectors and eigenvalues are determined numerically in a finite system by using a discretization procedure, the Goldstone modes will possess a small finite frequency. However, these will not be presented in Sec.~\ref{sec3a} together with the other frequencies since they represent a numerical artifact. For the $1\pi$ and $3\pi$ skyrmions the $\omega_{0,1}$ eigenmode has a positive frequency and an eigenvector clearly distinguishable from that of the $\omega_{0,-1}$ translational mode. However, for the  $2\pi$ skyrmion both the $\omega_{0,-1}$ and the $\omega_{0,1}$ eigenfrequencies of $\boldsymbol{D}_{\textrm{SW}}$ are very close to zero, and the corresponding eigenvectors converge to Eqs.~(\ref{eqn14}) and (\ref{eqn15}) as the discretization is refined and the system size is increased. This can occur because $\boldsymbol{D}_{\textrm{SW}}$ is not self-adjoint and its eigenvectors are generally not orthogonal. In contrast, the eigenvectors of $\boldsymbol{H}_{\textrm{SW}}$ remain orthogonal, with only a single pair of them taking the form of Eqs.~(\ref{eqn14}) and (\ref{eqn15}).


In contrast to the Goldstone modes with always zero energy, the sign change of another eigenvalue of $\boldsymbol{H}_{\textrm{SW}}$ indicates that the isolated $k\pi$ skyrmion is transformed from a stable local energy minimum into an unstable saddle point, leading to its disappearance from the system. Such instabilities were determined by calculating the lowest-lying eigenvalues of $\boldsymbol{H}_{\textrm{SW}}$ in Eq.~(\ref{eqn2}). Due to the connection between the $\boldsymbol{H}_{\textrm{SW}}$ and $\boldsymbol{D}_{\textrm{SW}}$ matrices expressed in Eq.~(\ref{eqn6}), at least one of the precession frequencies $\omega_{q}$ will also approach zero at such an instability point.

\subsection{Effective damping parameters\label{sec2e}}

For finite values of the Gilbert damping $\alpha$, the spin waves in the system will decay over time as the system relaxes to the equilibrium state during the time evolution described by the Landau--Lifshitz--Gilbert equation. The speed of the relaxation can be characterized by the effective damping parameter, which for a given mode $q$ is defined as
\begin{eqnarray}
\alpha_{q,\textrm{eff}}=\left|\frac{\textrm{Im}\:\omega_{q}}{\textrm{Re}\:\omega_{q}}\right|.\label{eqn7}
\end{eqnarray}

As discussed in detail in Ref.~\cite{Rozsa}, $\alpha_{q,\textrm{eff}}$ is mode-dependent and can be significantly higher than the Gilbert damping parameter $\alpha$ due to the elliptic polarization of spin waves, which can primarily be attributed to the noncollinear spin structure of the $k\pi$ skyrmions. For $\alpha\ll 1$, $\alpha_{q,\textrm{eff}}$ may be expressed as
\begin{eqnarray}
\frac{\alpha_{q,\textrm{eff}}}{\alpha}=\frac{\mathlarger{\sum}_{i}\left|\tilde{S}_{q,i}^{(0),x}\right|^{2}+\left|\tilde{S}_{q,i}^{(0),y}\right|^{2}}{\mathlarger{\sum}_{i}2\,\textrm{Im}\left[\left(\tilde{S}_{q,i}^{(0),x}\right)^{*}\tilde{S}_{q,i}^{(0),y}\right]},\label{eqn8}
\end{eqnarray}
where the eigenvectors in Eq.~(\ref{eqn8}) are calculated at $\alpha=0$ from Eq.~(\ref{eqn6}). Equation~(\ref{eqn8}) may also be expressed by the axes of the polarization ellipse of the spins in mode $q$, see Ref.~\cite{Rozsa} for details.

For higher values of $\alpha$, the complex frequencies $\omega_{q}$ have to be determined from Eq.~(\ref{eqn6}), while the effective damping parameters can be calculated from Eq.~(\ref{eqn7}). Also for finite values of $\alpha$ for each frequency with $\textrm{Re}\:\omega_{q}>0$ there exists a pair with $\textrm{Re}\:\omega_{q'}<0$ such that $\omega_{q'}=-\omega_{q}^{*}$ \cite{Rozsa}. The spin waves will be circularly polarized if $\boldsymbol{A}_{1}=\boldsymbol{A}_{3}$ and $\boldsymbol{A}_{2}^{\dag}=-\boldsymbol{A}_{2}$ in Eq.~(\ref{eqn2}), in which case the dependence of $\omega_{q}$ on $\alpha$ may simply be expressed by the undamped frequency $\omega_{q}^{\left(0\right)}$ as
\begin{eqnarray}
\textrm{Re}\:\omega_{q}\left(\alpha\right)=\frac{1}{1+\alpha^{2}}\omega_{q}^{\left(0\right)},\label{eqn8a}
\\
\left|\textrm{Im}\:\omega_{q}\left(\alpha\right)\right|=\frac{\alpha}{1+\alpha^{2}}\omega_{q}^{\left(0\right)}.\label{eqn8b}
\end{eqnarray}

These relations are known for uniaxial ferromagnets; see, e.g., Ref.~\cite{Rozsa2}. In the elliptically polarized modes of noncollinear structures, such as $k\pi$ skyrmions, a deviation from Eqs.~(\ref{eqn8a})-(\ref{eqn8b}) is expected.

\section{Results}

\subsection{Eigenmodes\label{sec3a}}

\begin{figure}
\centering
\includegraphics[width=\columnwidth]{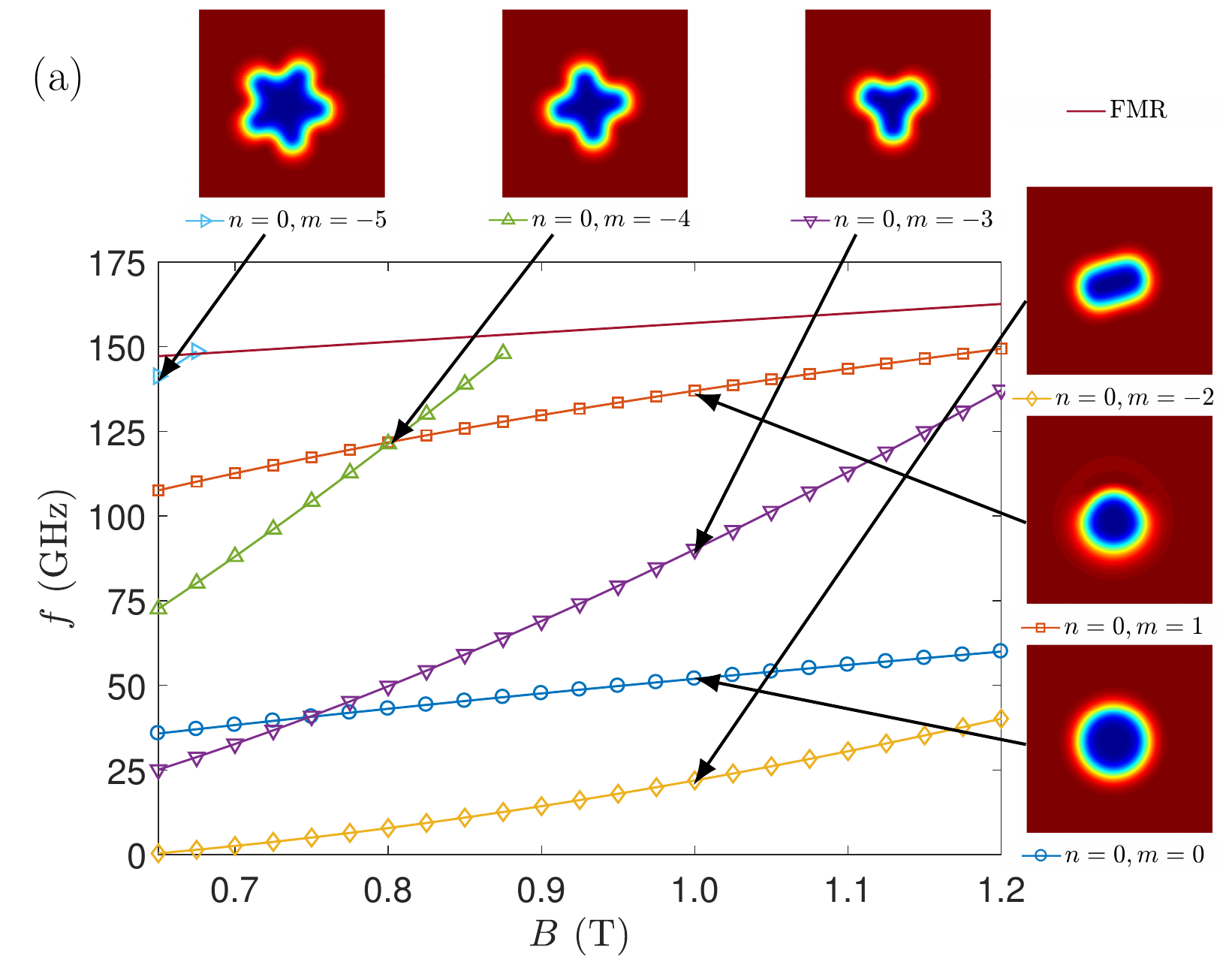}
\includegraphics[width=\columnwidth]{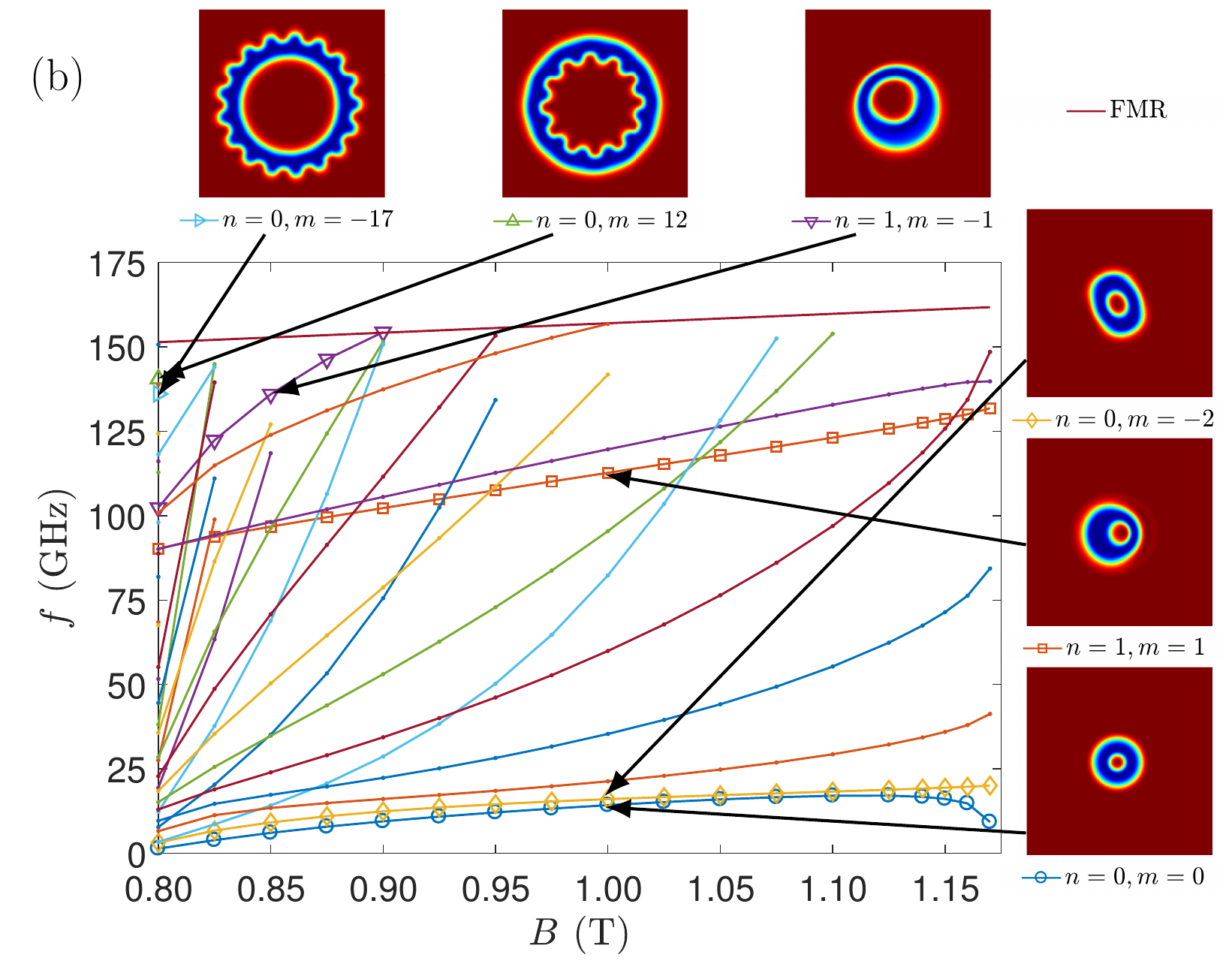}
\includegraphics[width=\columnwidth]{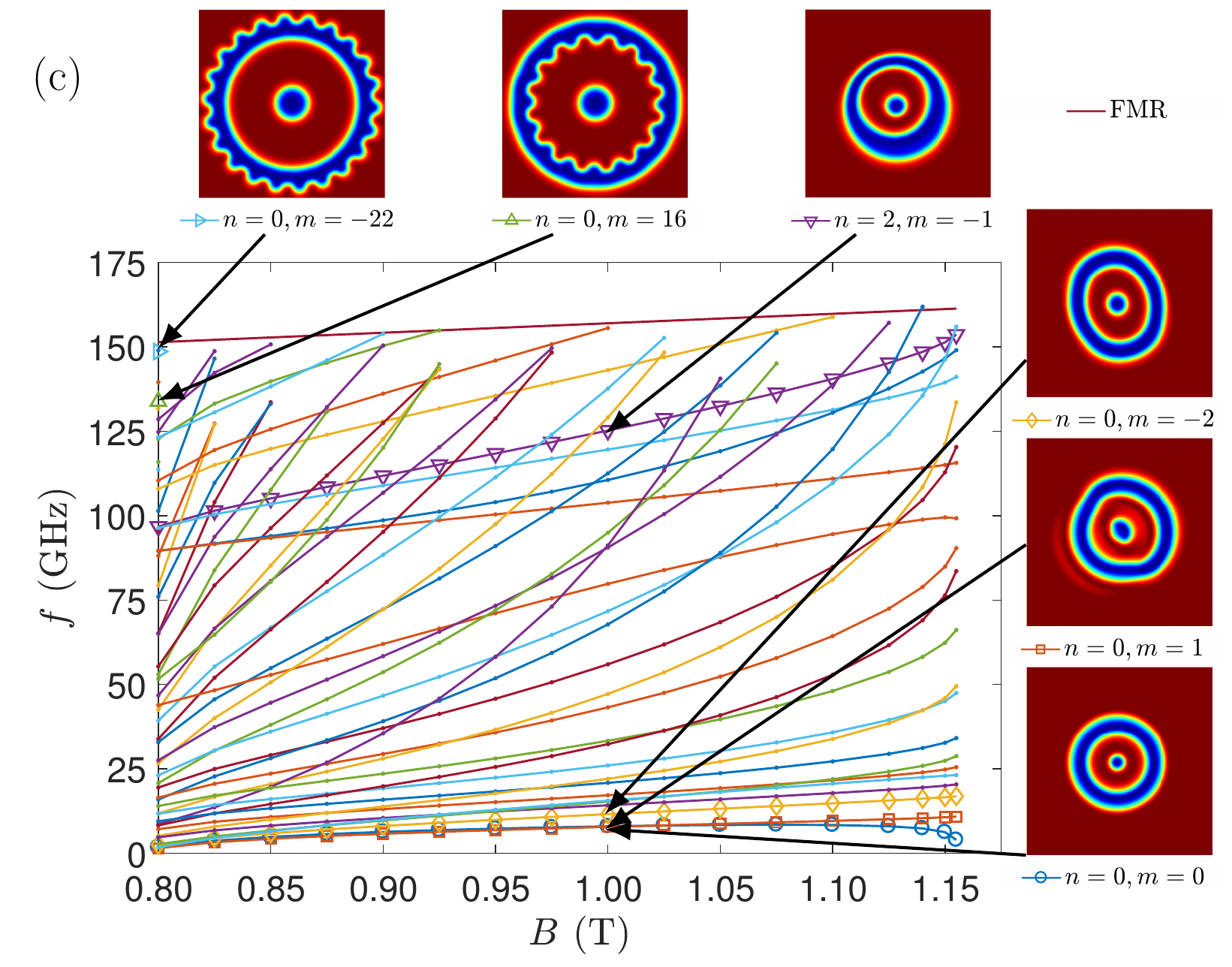}
\caption{Frequencies of localized spin wave modes at $\alpha=0$ for (a) the $1\pi$, (b) the $2\pi$, and (c) the $3\pi$ skyrmion. Selected spin wave modes are
visualized in contour plots of the out-of-plane spin component and 
denoted by open symbols connected by lines in the figure, the remaining modes are denoted by connected dots.\label{fig1}}
\end{figure}

The frequencies of the localized spin wave modes of the $1\pi$, $2\pi$, and $3\pi$ skyrmion, calculated from the atomistic model for $\alpha=0$ as described in Sec.~\ref{sec2a}, are shown in Fig.~\ref{fig1}. For the $1\pi$ skyrmion six localized modes can be observed below the FMR frequency of the field-polarized background in Fig.~\ref{fig1}(a), four of which are clockwise rotating modes ($m<0$), one is a gyration mode rotating counterclockwise ($m=1$), while the final one is a breathing mode ($m=0$). The excitation frequencies show good quantitative agreement with the ones calculated from the micromagnetic model for the same system in Ref.~\cite{Rozsa}. Compared to Ref.~\cite{Schutte}, the additional appearance of the eigenmodes with $m=1,-4,-5$  can be attributed to the finite value of the anisotropy parameter $K$ in the present case. Increasing the anisotropy value makes it possible to stabilize the skyrmions at lower field values, down to zero field at the critical value in the micromagnetic model $\left|\mathcal{K}_{\textrm{c}}\right|=\dfrac{\pi^{2}\mathcal{D}^{2}}{16\mathcal{A}}$, where the transition from the spin spiral to the ferromagnetic ground state occurs at zero external field \cite{Bogdanov3}. Since the excitation frequencies decrease at lower field values as shown in Fig.~\ref{fig1}(a), this favors the appearance of further modes. Simultaneously, the FMR frequency increases with $K$, meaning that modes with higher frequencies become observable for larger uniaxial anisotropy. For each angular momentum quantum number $m$, only a single mode ($n=0$) appears.

In the case of the $2\pi$ skyrmion an increased number of eigenmodes may be seen in Fig.~\ref{fig1}(b). This can mainly be attributed to the appearance of spin waves with higher angular momentum quantum numbers both for clockwise (up to $m=-17$) and counterclockwise (up to $m=12$) rotational directions. Furthermore, in this case modes with $n=1$ node in the eigenfunction can be observed as well. The same trend continues in the case of $3\pi$ skyrmions in Fig.~\ref{fig1}(c), the large number of internal eigenmodes can be attributed to angular momentum quantum numbers ranging from $m=-22$ to $m=16$, as well as to spin wave eigenvectors with up to $n=2$ nodes. The different rotational directions and numbers of nodes are illustrated in Supplemental Videos 1-4 \cite{supp} via the square-shaped modes ($n=0,1$, $m=\pm 4$) of the $3\pi$ skyrmion at $B=0.825\,\textrm{T}$.

The increase of possible angular momentum quantum numbers for higher skyrmion order $k$ as well as for decreasing magnetic field $B$ may be qualitatively explained by an increase in the skyrmion size. Modes with a given value of $m$ indicate a total of $\left|m\right|$ modulation periods along the perimeter of the skyrmion; for larger skyrmion sizes this corresponds to a modulation on a longer length scale, which has a smaller cost in exchange energy.

\begin{figure}
\centering
\includegraphics[width=\columnwidth]{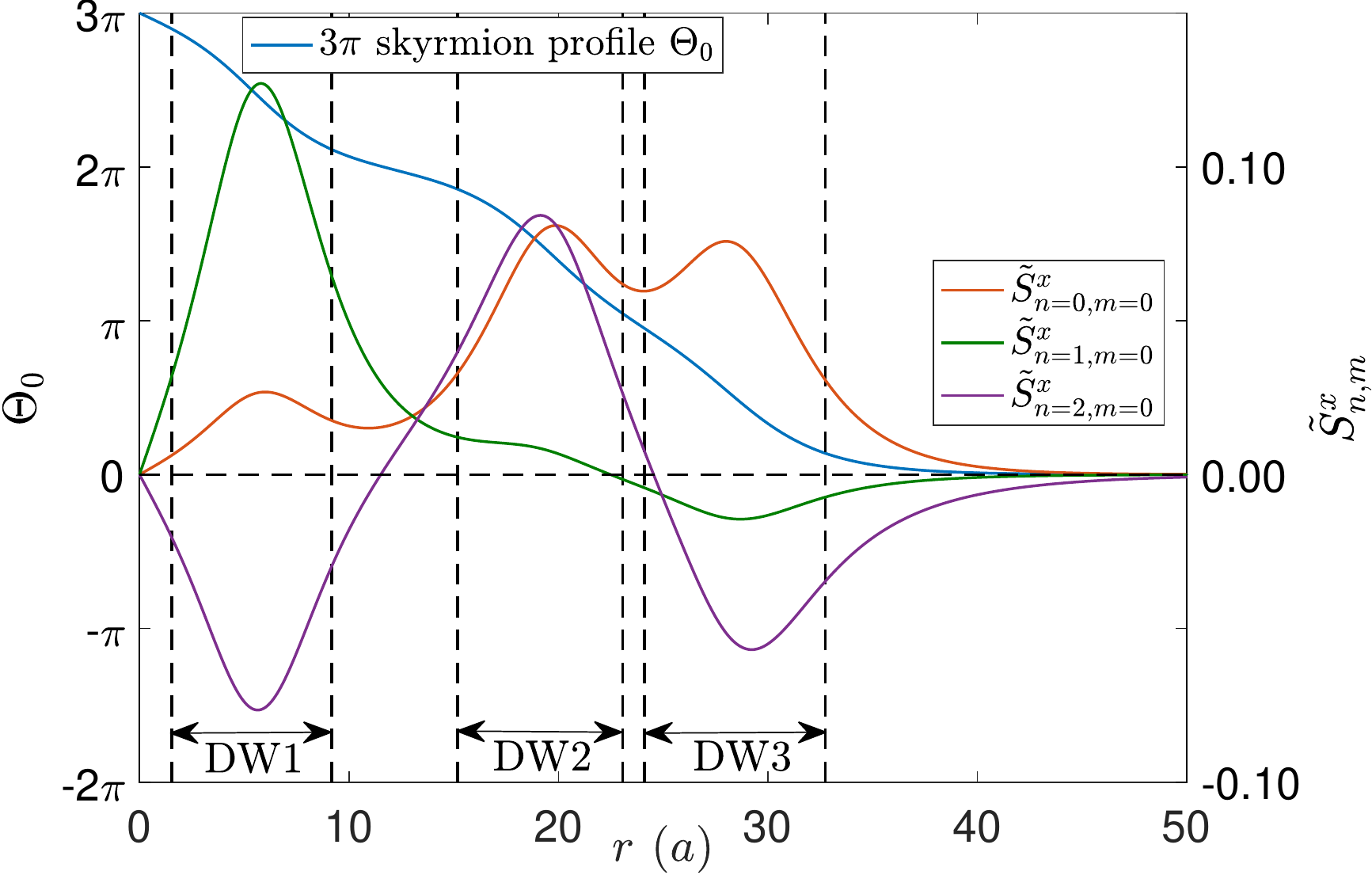}
\caption{Comparison between the $3\pi$ skyrmion profile (left vertical axis) and the eigenvectors of the breathing modes ($m=0$) with different numbers of nodes $n=0,1,2$ (right vertical axis). The calculations were performed using the micromagnetic model described in Sec.~\ref{sec2b} at $B=1\,\textrm{T}$, the lattice constant is $a=0.271\,\textrm{nm}$. Double arrows between vertical dashed lines indicate the extensions of the domain walls in the structure.\label{fig2}}
\end{figure}

The breathing modes of the $3\pi$ skyrmion with different numbers of nodes are visualized in Fig.~\ref{fig2} at $B=1\,\textrm{T}$. The results shown in Fig.~\ref{fig2} are obtained from the micromagnetic model in Sec.~\ref{sec2b}, which is in good quantitative agreement with the atomistic calculations at the given field. All the eigenmodes display three peaks of various heights, while they decay exponentially outside the $3\pi$ skyrmion. As can be seen in Fig.~\ref{fig2}, the peaks are localized roughly around the regions where the spins are lying in-plane, indicated by the domain walls (DW) between pairs of dashed lines. The widths of the domain walls were determined by approximating the $3\pi$ skyrmion profile with linear functions close to the inflection points $r_{j},\Theta_{0,j},j=1,2,3$ where the spins are lying in-plane, and calculating where these linear functions intersect integer multiples of $\pi$ in $\Theta_{0}$. Thus, the domain walls are located between the inner $R_{\textrm{in},j}=r_{j}+\left[\partial_{r}\Theta_{0}\left(r_{j}\right)\right]^{-1}\left[\left(4-j\right)\pi-\Theta_{0,j}\right]$ and outer $R_{\textrm{out},j}=r_{j}+\left[\partial_{r}\Theta_{0}\left(r_{j}\right)\right]^{-1}\left[\left(3-j\right)\pi-\Theta_{0,j}\right]$ radii. Such a description was used to calculate the skyrmion radius in, e.g., Ref.~\cite{Bogdanov3}, and it was also applied for calculating the widths of planar domain walls \cite{Hubert}.

The nodes of the eigenmodes are located roughly between these domain walls, meaning that typically excitation modes with $n=0,\dots,k-1$ nodes may be observed in $k\pi$ skyrmions, in agreement with the results in Fig.~\ref{fig1}. A higher number of nodes would require splitting a single peak into multiple peaks, the energy cost of which generally exceeds the FMR frequency, thereby making these modes unobservable. The sign changes in the $\tilde{S}^{x}_{n,m}$ eigenvectors mean that the different modes can be imagined as the domain walls breathing in the same phase or in opposite phase, as can be seen in Supplemental Videos 5-7 \cite{supp}. Note that eigenmodes with higher $n$ quantum numbers may also be observed for skyrmions confined in nanodots \cite{Kim,Gareeva,Beg} where the peaks of the eigenmodes may also be localized at the edge of the sample, in contrast to the present case where isolated $k\pi$ skyrmions are discussed on an infinite collinear background.

It is also worth noting that the lowest-lying nonzero gyration mode is $n=0,m=1$ for the $1\pi$ and $3\pi$ skyrmions, while it is $n=1,m=1$ for the $2\pi$ skyrmion, see Fig.~\ref{fig1}. As already mentioned in Sec.~\ref{sec2c}, numerical calculations for the $2\pi$ skyrmion indicate both in the atomistic and the micromagnetic case that by increasing the system size or refining the discretization the eigenvectors of both the $n=0,m=-1$ and the $n=0,m=1$ modes of $\boldsymbol{D}_{\textrm{SW}}$ in Eq.~(\ref{eqn6}) converge to the same eigenvectors in Eqs.~(\ref{eqn14}) and (\ref{eqn15}) and $0$ eigenvalue, which correspond to the translational Goldstone mode in the infinite system. 
This difference can probably be attributed to the deviation in the value of the topological charge, being finite for $1\pi$ and $3\pi$ skyrmions but zero for the $2\pi$ skyrmion \cite{Hagemeister}.

\subsection{Instabilities\label{sec3b}}

Skyrmions with different order $k$ deviate in their low-field behavior. Since the considered Pd/Fe/Ir(111) system has a spin spiral ground state \cite{Romming}, decreasing the magnetic field value will make the formation of domain walls energetically preferable in the system. In the case of the $1\pi$ skyrmion this means that the lowest-lying eigenmode of $\boldsymbol{H}_{\textrm{SW}}$ in Eq.~(\ref{eqn2}), which is an elliptic mode with $m=\pm 2$, changes sign from positive to negative, occurring between $B=0.650\,\textrm{T}$ and $B=0.625\,\textrm{T}$ in the present system. This is indicated in Fig.~\ref{fig1}(a) by the fact that the frequency of the $n=0,m=-2$ eigenmode of $\boldsymbol{D}_{\textrm{SW}}$ in Eq.~(\ref{eqn6}) converges to zero. This leads to an elongation of the skyrmion into a spin spiral segment which gradually fills the ferromagnetic background, a so-called strip-out or elliptic instability already discussed in previous publications \cite{Bogdanov3,Schutte}. In contrast, for the $2\pi$ and $3\pi$ skyrmions the lowest-lying eigenmode of $\boldsymbol{H}_{\textrm{SW}}$ is a breathing mode with $m=0$, which tends to zero between $B=0.800\,\textrm{T}$ and $B=0.775\,\textrm{T}$ for both skyrmions. This is indicated by the lowest-lying $n=0,m=0$ mode of $\boldsymbol{D}_{\textrm{SW}}$ in Fig.~\ref{fig1}(b) for the $2\pi$ skyrmion, which is the second lowest after the $n=0,m=1$ mode for the $3\pi$ skyrmion in Fig.~\ref{fig1}(c). This means that the radius of the outer two rings of $2\pi$ and $3\pi$ skyrmions diverges at a finite field value, leading to a burst instability. Such a type of instability was already shown to occur in Ref.~\cite{Bogdanov4} in the case of a ferromagnetic ground state at negative field values, in which case it also affects $1\pi$ skyrmions.

At the burst instability, modes with $n=0$ and all angular momentum quantum numbers $m$ appear to approach zero because of the drastic increase in skyrmion radius decreasing the frequency of these modes as discussed in Sec.~\ref{sec3a}. 
A similar effect was observed for the $1\pi$ skyrmion in Ref.~\cite{Kravchuk} when the critical value of the Dzyaloshinsky--Moriya interaction, $\left|\mathcal{D}_{c}\right|=\frac{4}{\pi}\sqrt{\mathcal{A}\left|\mathcal{K}\right|}$, was approached at zero external field from the direction of the ferromagnetic ground state. 
In contrast, the elliptic instability only seems to affect the $n=0,m=-2$ mode, while other $m$ values and the nonreciprocity are apparently weakly influenced.

\begin{figure}
\centering
\includegraphics[width=\columnwidth]{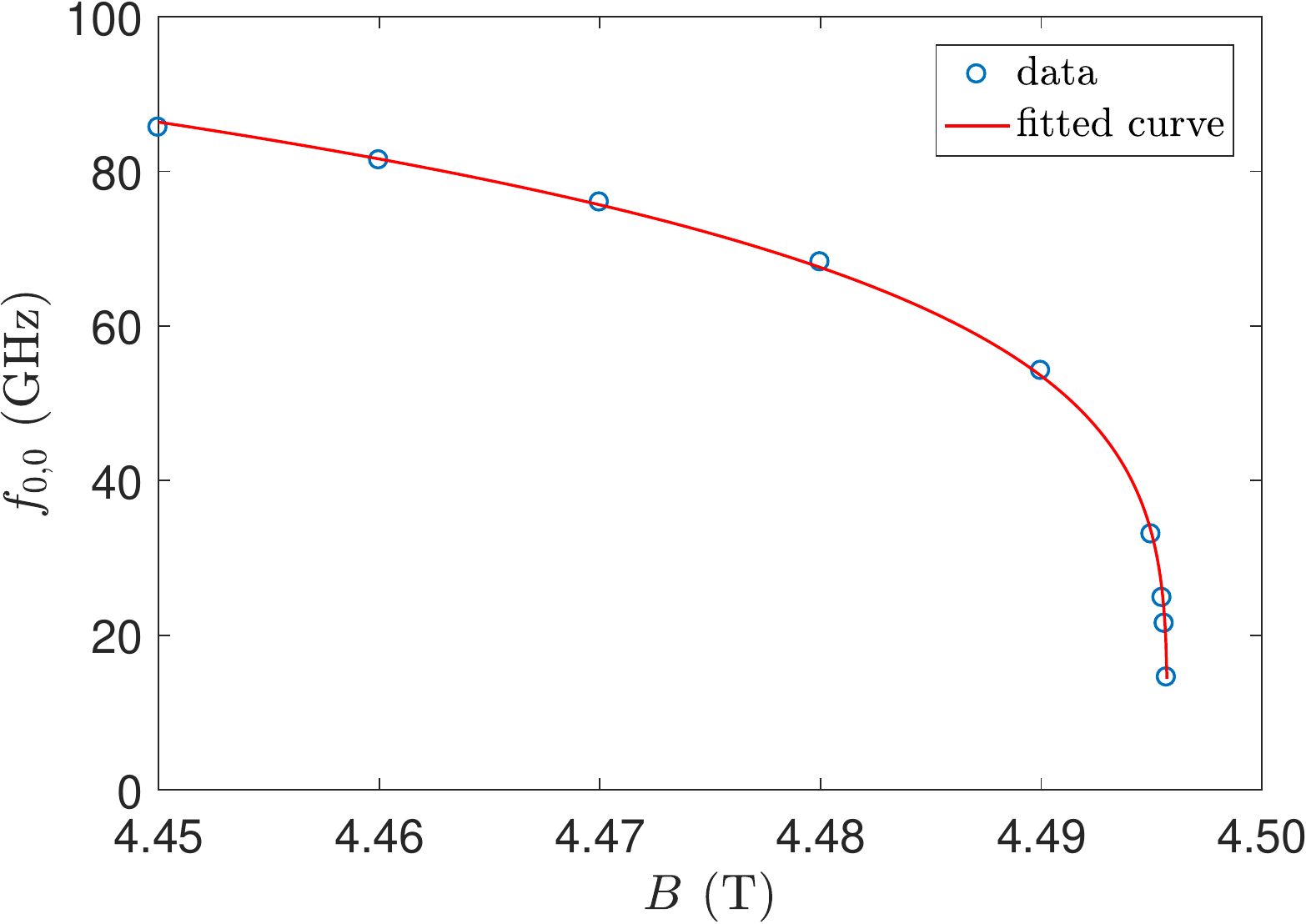}
\caption{Frequency of the breathing mode $n=0,m=0$ of the $1\pi$ skyrmion close to the collapse field. Calculation data are shown by open symbols, red line denotes the power-law fit $f_{0,0}=A_{f}\left(B_{\textrm{c},1\pi}-B\right)^{\beta_{f}}$.\label{fig3}}
\end{figure}

In the atomistic model, skyrmions collapse when their characteristic size becomes comparable to the lattice constant. For the $1\pi$, $2\pi$, and $3\pi$ skyrmions the collapse of the innermost ring occurs at  $B_{\textrm{c},1\pi}\approx4.495\,\textrm{T}$, $B_{\textrm{c},2\pi}\approx1.175\,\textrm{T}$, and $B_{\textrm{c},3\pi}\approx1.155\,\textrm{T}$, respectively \cite{Hagemeister}. As can be seen in Figs.~\ref{fig1}(b), \ref{fig1}(c), and \ref{fig3}, this instability is again signaled by the $n=0,m=0$ eigenfrequency going to zero, but in contrast to the burst instability, the other excitation frequencies keep increasing with the field in this regime. Figure~\ref{fig3} demonstrates that close to the collapse field the excitation frequency may be well approximated by the power law $f_{0,0}=A_{f}\left(B_{\textrm{c},1\pi}-B\right)^{\beta_{f}}$, with $A_{f}=175.6\,\frac{\textrm{GHz}}{\textrm{T}^{\beta_{f}}}$, $B_{\textrm{c},1\pi}=4.4957\,\textrm{T}$, and $\beta_{f}=0.23$. 

\subsection{Effective damping parameters in the limit of low $\alpha$\label{sec3c}}

\begin{figure}
\centering
\includegraphics[width=0.98\columnwidth]{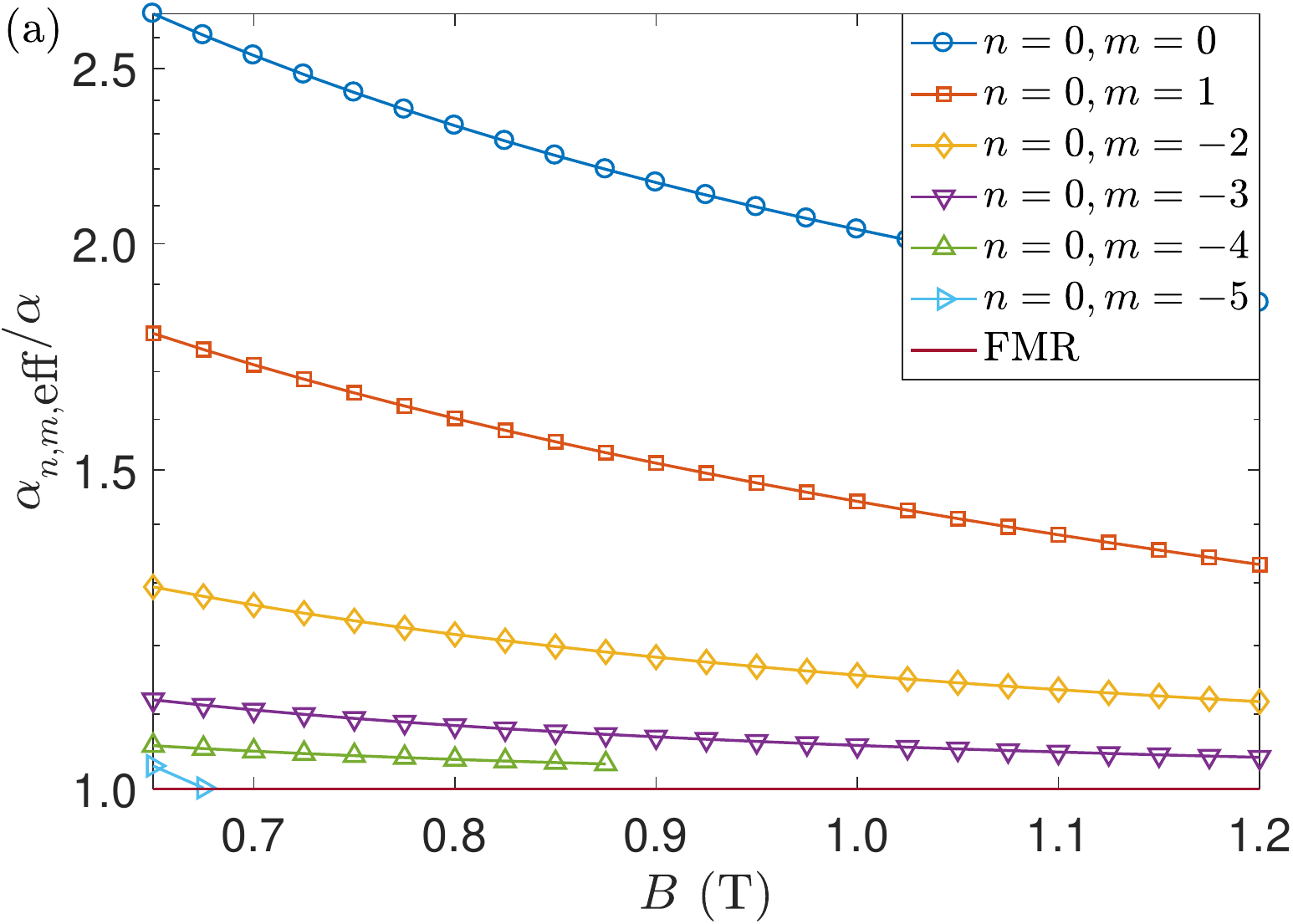}
\includegraphics[width=0.98\columnwidth]{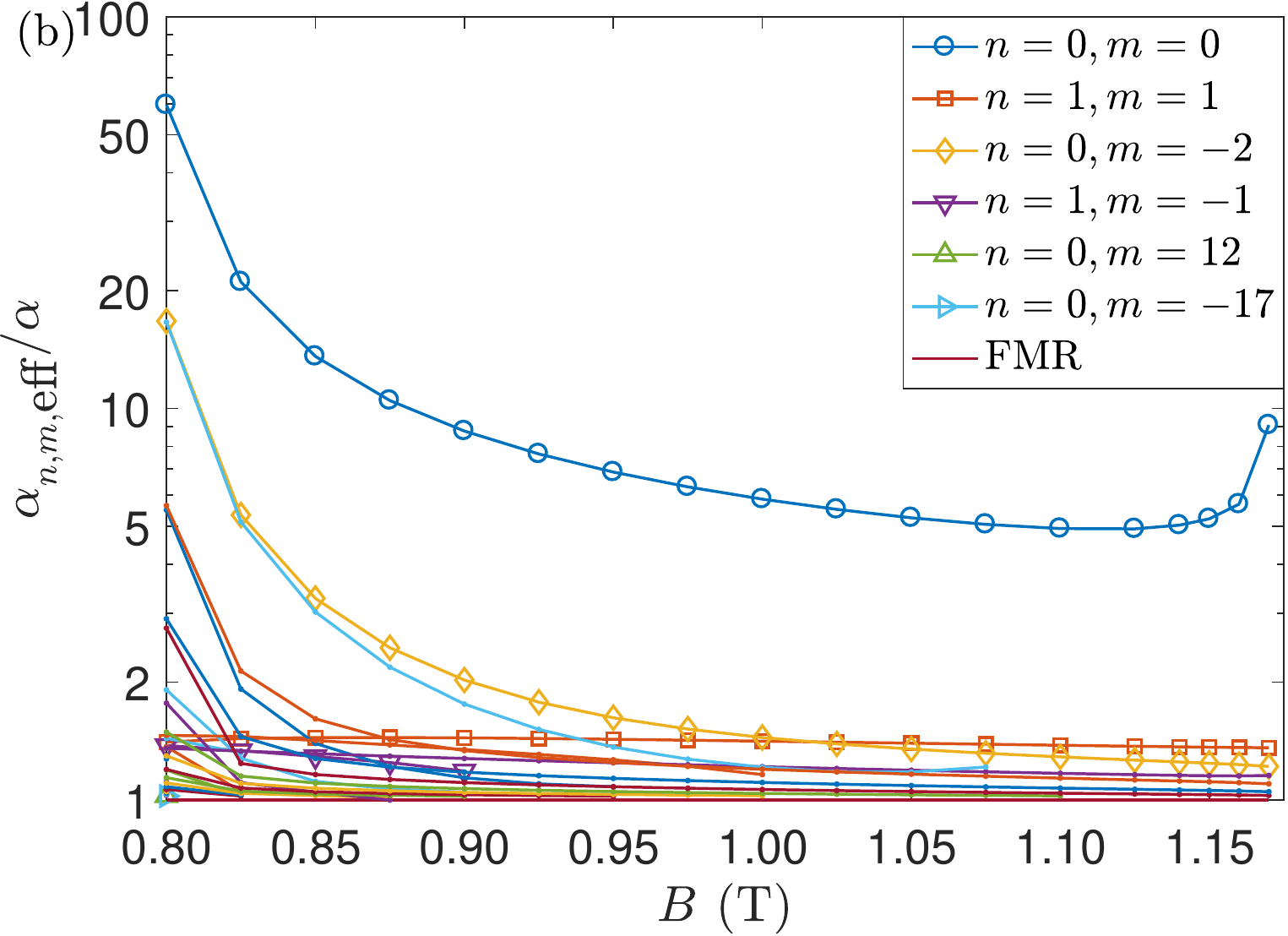}
\includegraphics[width=0.98\columnwidth]{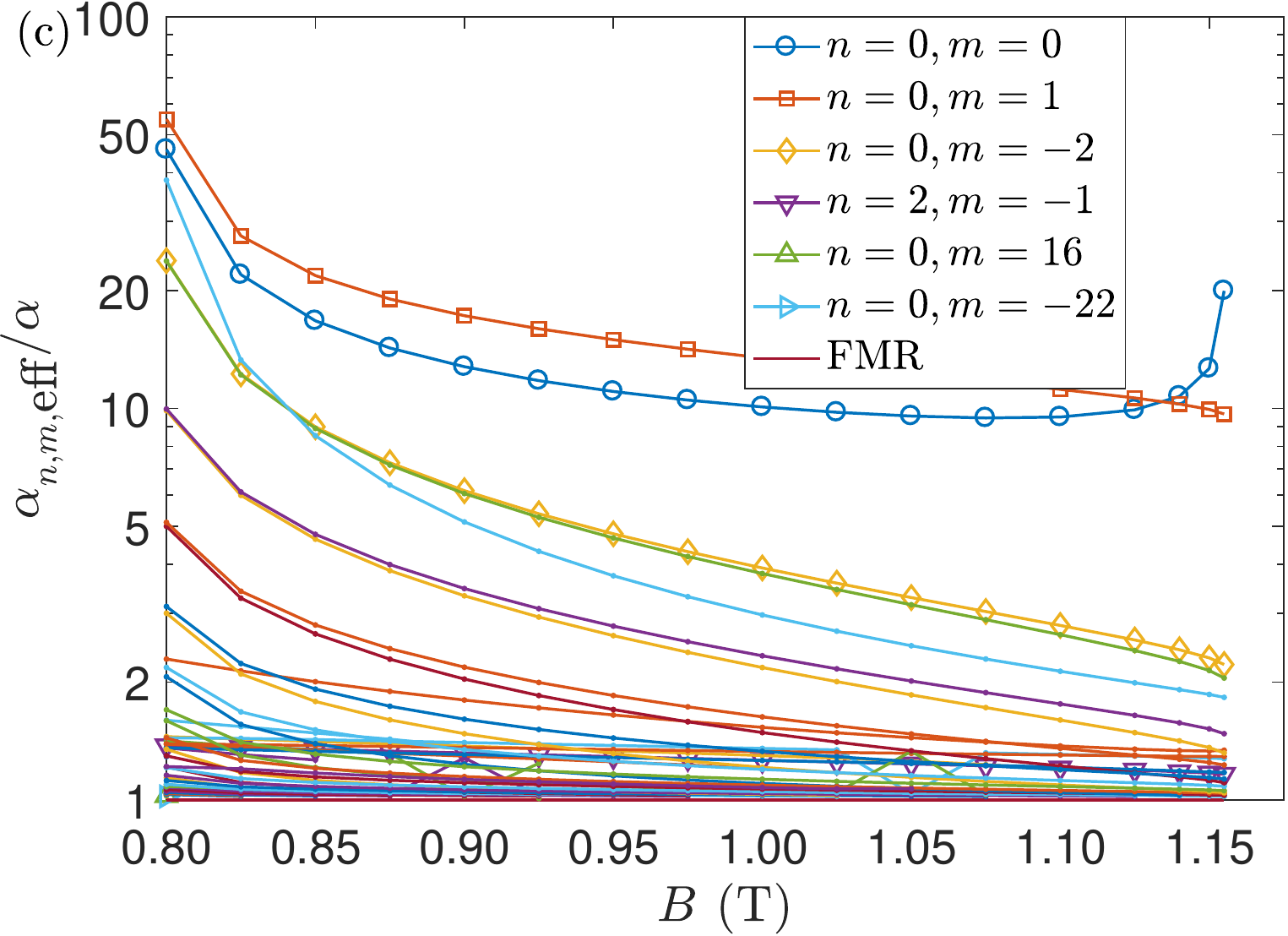}
\caption{Effective damping parameters calculated according to Eq.~(\ref{eqn8}) for the eigenmodes of the (a) $1\pi$, (b) $2\pi$, and (c) $3\pi$ skyrmions, plotted on a logarithmic scale. The corresponding excitation frequencies are shown in Fig.~\ref{fig1}.\label{fig4}}
\end{figure}

The effective damping parameters $\alpha_{n,m,\textrm{eff}}$ were first calculated from the eigenvectors obtained at $\alpha=0$ following Eq.~(\ref{eqn8}). The results for the $1\pi$, $2\pi$, and $3\pi$ skyrmions are summarized in Fig.~\ref{fig4}. As discussed in Ref.~\cite{Rozsa}, the $\alpha_{n,m,\textrm{eff}}$ values are always larger than the Gilbert damping $\alpha$, and they tend to decrease with increasing angular momentum quantum number $\left|m\right|$ and magnetic field $B$. The spin wave possessing the highest effective damping is the $n=0,m=0$ breathing mode both for the $1\pi$ and $2\pi$ skyrmion, but it is the $n=0,m=1$ gyration mode for the $3\pi$ skyrmion for a large part of the external field range where the structure is stable. Excitation pairs with quantum numbers $n,\pm m$ tend to decay with similar $\alpha_{n,m,\textrm{eff}}$ values to each other, with $\alpha_{n,\left|m\right|,\textrm{eff}}<\alpha_{n,-\left|m\right|,\textrm{eff}}$, where clockwise modes ($m<0$) have lower frequencies and higher effective damping due to the nonreciprocity.

The effective damping parameters drastically increase and for the lowest-lying modes apparently diverge close to the burst instability, while no such sign of nonanalytical behavior can be observed in the case of the $1\pi$ skyrmion with the elliptic instability.
For the same $n,m$ mode, the effective damping parameter tends to increase with skyrmion order $k$ away from the critical field regimes; for example, for the $n=0,m=0$ mode at $B=1.00\,\textrm{T}$ one finds $\alpha_{0,0,\textrm{eff},1\pi}=2.04$, $\alpha_{0,0,\textrm{eff},2\pi}=5.87$, and $\alpha_{0,0,\textrm{eff},3\pi}=10.09$.

\begin{figure}
\centering
\includegraphics[width=\columnwidth]{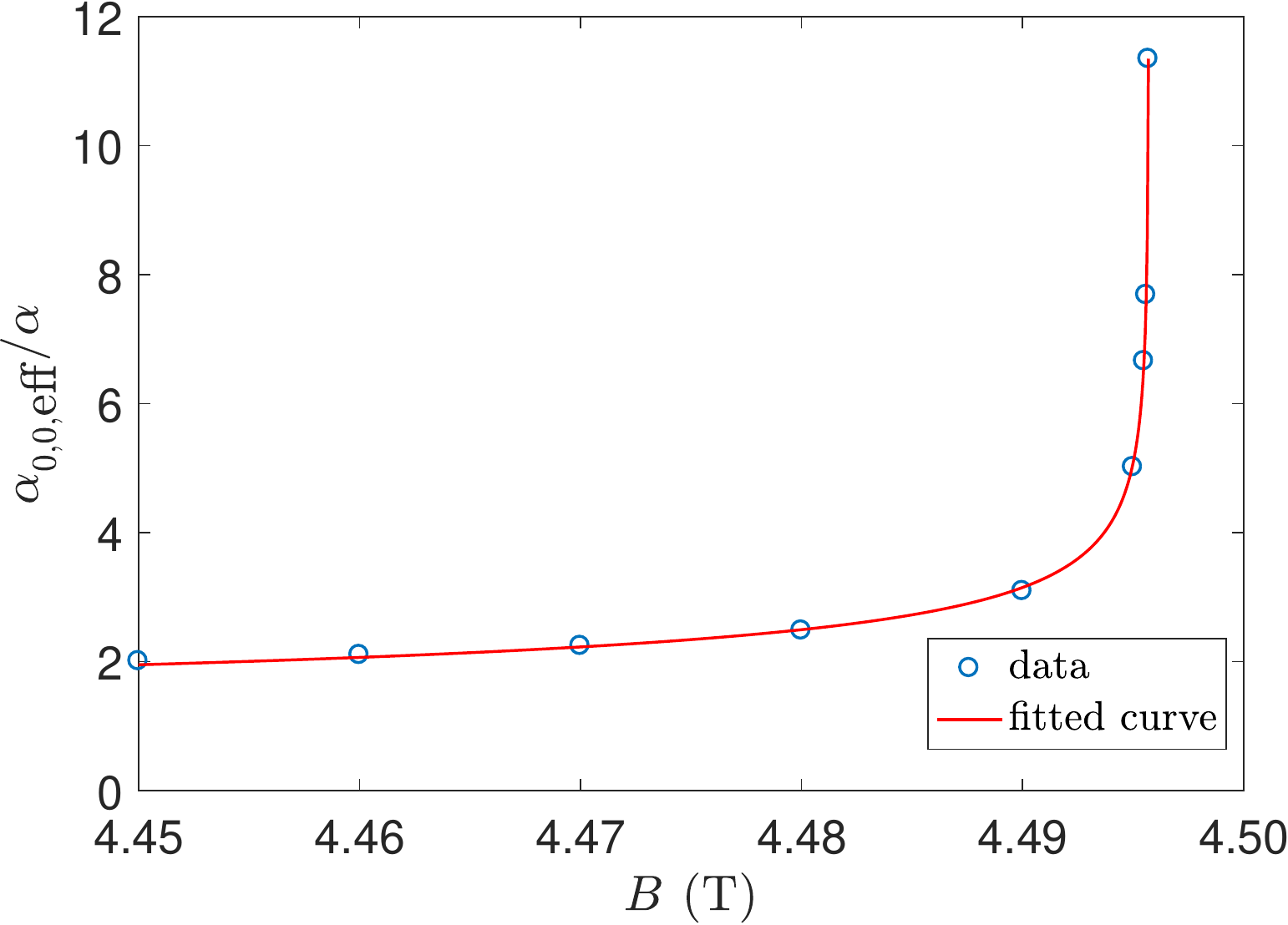}
\caption{Effective damping parameter $\alpha_{0,0,\textrm{eff}}$ of the breathing mode $n=0,m=0$ of the $1\pi$ skyrmion close to the collapse field. The corresponding excitation frequencies are shown in Fig.~\ref{fig3}. Calculation data are shown by open symbols, red line denotes the power-law fit $\alpha_{0,0,\textrm{eff}}=A_{\alpha}\left(B_{\textrm{c},1\pi}-B\right)^{-\beta_{\alpha}}$.\label{fig5}}
\end{figure}

Close to the collapse field, the effective damping parameter of the $n=0,m=0$ breathing mode tends to diverge as shown in Figs.~\ref{fig4}(b), \ref{fig4}(c), and \ref{fig5} for the $2\pi$, $3\pi$, and $1\pi$ skyrmions, respectively. Similarly to the eigenfrequency converging to zero in Fig.~\ref{fig3}, the critical behavior of the effective damping may be approximated by a power-law fit $\alpha_{0,0,\textrm{eff}}=A_{\alpha}\left(B_{\textrm{c},1\pi}-B\right)^{-\beta_{\alpha}}$ as shown in Fig.~\ref{fig5}, this time with a negative exponent due to the divergence. The fitting yields the parameters $A_{\alpha}=0.96\,\textrm{T}^{\beta_{\alpha}}$, $B_{\textrm{c},1\pi}=4.4957\,\textrm{T}$, and $\beta_{\alpha}=0.23$. Naturally, the critical field values agree between the two fits, but interestingly one also finds $\beta_{f}=\beta_{\alpha}$ up to two digits precision. Rearranging Eq.~(\ref{eqn7}) yields
\begin{eqnarray}
\frac{\alpha_{0,0,\textrm{eff}}}{\alpha}\textrm{Re}\:\omega_{0,0}=\frac{1}{\alpha}\left|\textrm{Im}\:\omega_{0,0}\right|,\label{eqn16}
\end{eqnarray}
where the left-hand side is proportional to $\left(B_{\textrm{c},1\pi}-B\right)^{\beta_{f}-\beta_{\alpha}}$ which is approximately constant due to the exponents canceling. This indicates that while $\textrm{Re}\:\omega_{0,0}$ diverges close to the collapse field, $\left|\textrm{Im}\:\omega_{0,0}\right|/\alpha$ remains almost constant at low $\alpha$ values.

\subsection{Damping for higher $\alpha$ values\label{sec3d}}

\begin{figure}
\centering
\includegraphics[width=\columnwidth]{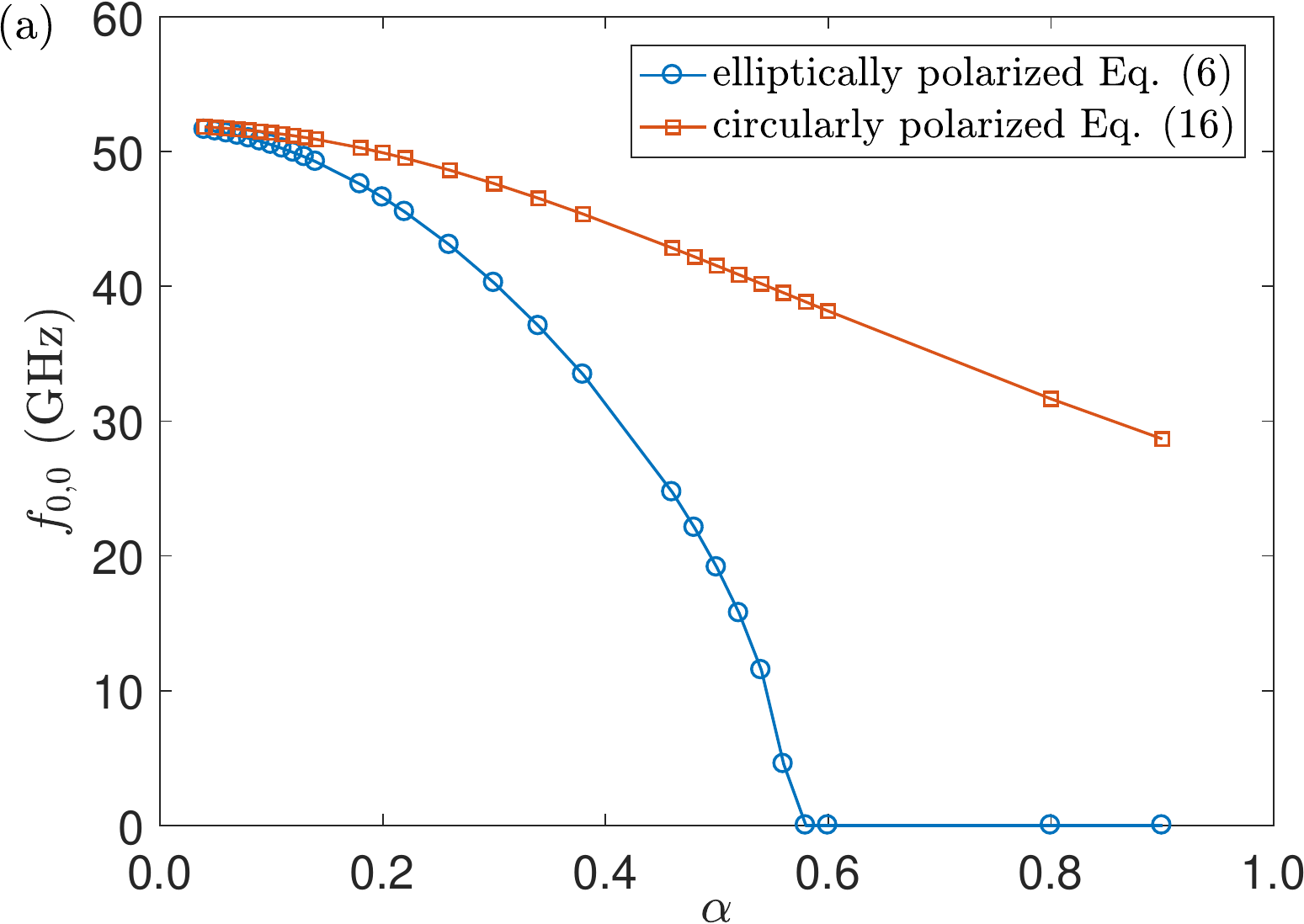}
\includegraphics[width=\columnwidth]{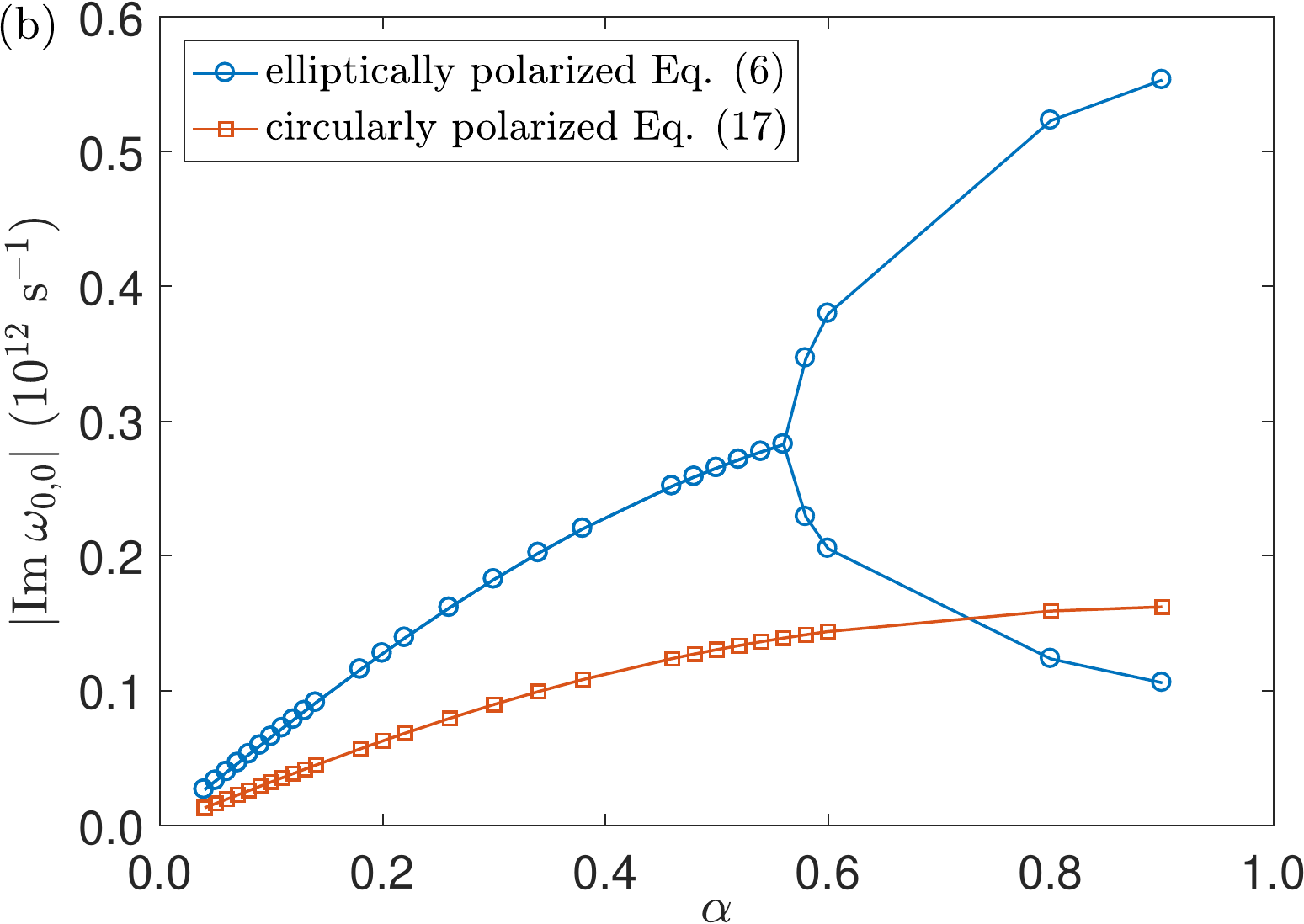}
\caption{(a) Frequency $f_{0,0}=\textrm{Re}\:\omega_{0,0}/2\pi$ and (b) inverse lifetime $\left|\textrm{Im}\:\omega_{0,0}\right|$ of the $n=0,m=0$ breathing mode of the $1\pi$ skyrmion at $B=1\,\textrm{T}$ as a function of the Gilbert damping parameter $\alpha$. The solutions of Eq.~(\ref{eqn6}) for the elliptically polarized eigenmode of the $1\pi$ skyrmion are compared to Eqs.~(\ref{eqn8a})-(\ref{eqn8b}) which are only valid for circularly polarized modes.\label{fig6}}
\end{figure}

Due to the divergences of the effective damping parameters found at the burst instability and collapse fields, it is worthwhile to investigate the consequences of using a finite $\alpha$ value in Eq.~(\ref{eqn6}), in contrast to relying on Eq.~(\ref{eqn8}) which is determined from the eigenvectors at $\alpha=0$. The $\alpha$ dependence of the real and imaginary parts of the $\omega_{0,0}$ breathing mode frequency of the $1\pi$ skyrmion is displayed in Fig.~\ref{fig6}, at a field value of $B=1\,\textrm{T}$ far from the elliptic and collapse instabilities. As shown in Fig.~\ref{fig6}(a), unlike circularly polarized modes described by Eq.~(\ref{eqn8a}) where $\textrm{Re}\:\omega_{q}$ decreases smoothly and equals half of the undamped value at $\alpha=1$, the $\textrm{Re}\:\omega_{0,0}$ value for the elliptically polarized eigenmode displays a much faster decay and reaches exactly zero at around $\alpha\approx 0.58$. According to Eq.~(\ref{eqn7}), this indicates that the corresponding effective damping parameter $\alpha_{0,0,\textrm{eff}}$ diverges at this point.

\begin{figure}
\centering
\includegraphics[width=\columnwidth]{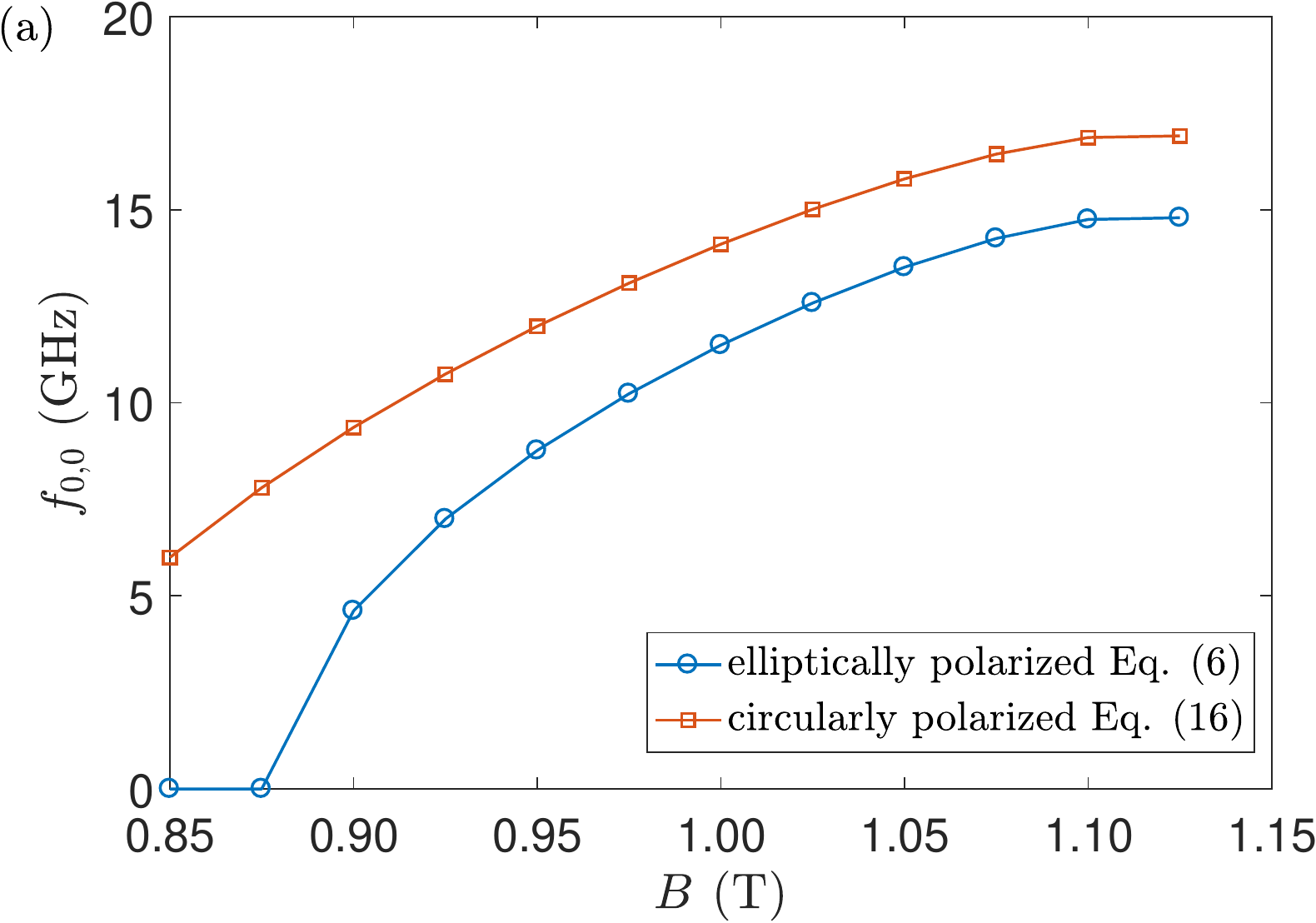}
\includegraphics[width=\columnwidth]{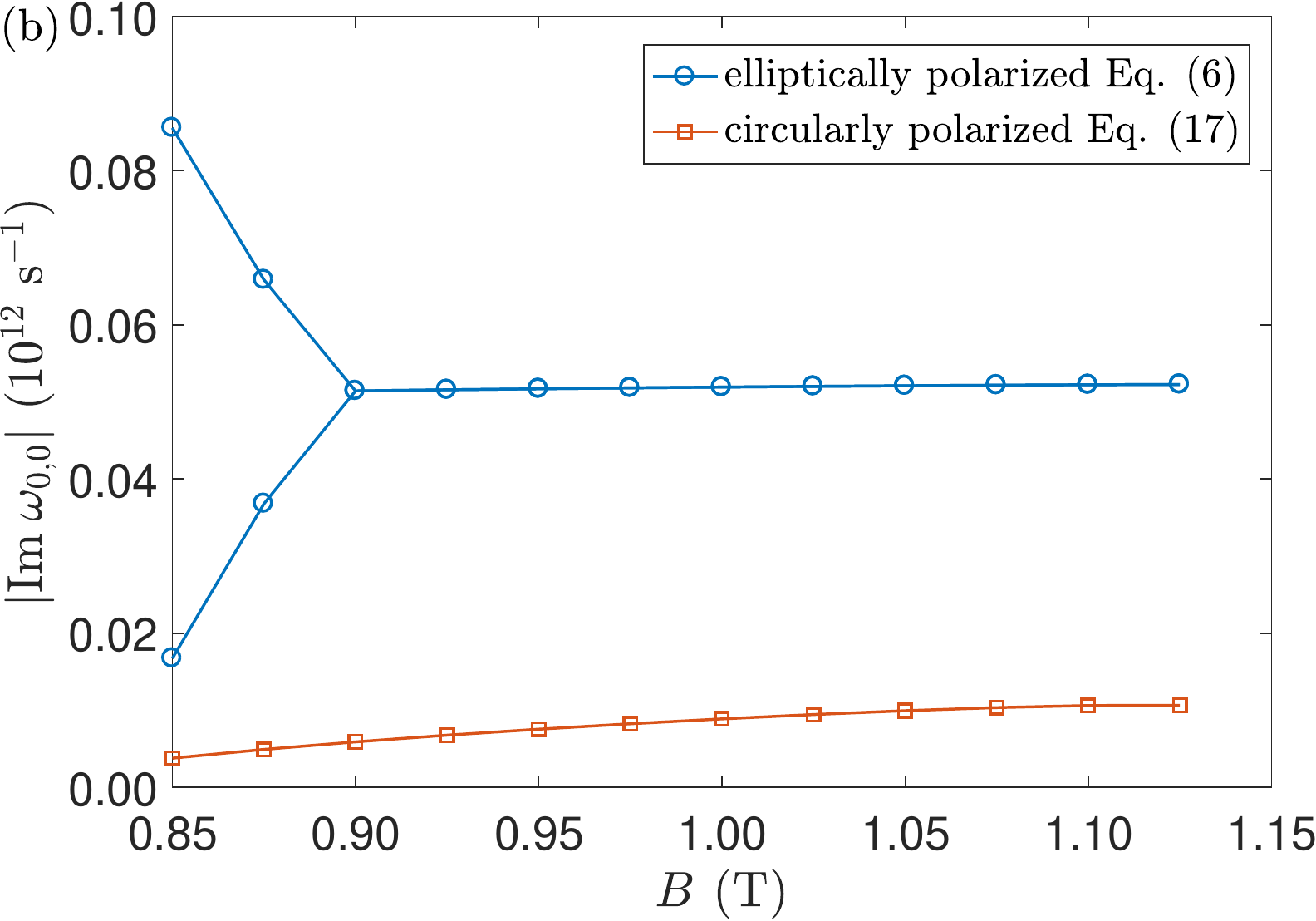}
\caption{(a) Frequency $f_{0,0}=\textrm{Re}\:\omega_{0,0}/2\pi$ and (b) inverse lifetime $\left|\textrm{Im}\:\omega_{0,0}\right|$ of the $n=0,m=0$ breathing mode of the $2\pi$ skyrmion at $\alpha=0.1$ as a function of the external magnetic field $B$. The solutions of Eq.~(\ref{eqn6}) for the elliptically polarized eigenmode of the $2\pi$ skyrmion are compared to Eqs.~(\ref{eqn8a})-(\ref{eqn8b}) which are only valid for circularly polarized modes.\label{fig7}}
\end{figure}

Since the real part of the frequency disappears, the $\omega_{q'}=-\omega_{q}^{*}$ relation connecting $\textrm{Re}\:\omega_{q}>0$ and $\textrm{Re}\:\omega_{q'}<0$ solutions of Eq.~(\ref{eqn6}) discussed in Sec.~\ref{sec2e} no longer holds, and two different purely imaginary eigenfrequencies are found in this regime as shown in Fig.~\ref{fig6}(b). This is analogous to overdamping in a classical linear harmonic oscillator, meaning that the purely precessional first-order differential equation describing circularly polarized modes is transformed into two coupled first-order differential equations \cite{Rozsa} with an effective mass term for the breathing mode of $k\pi$ skyrmions. 
This implies that when performing spin dynamics simulations based on the Landau--Lifshitz--Gilbert equation, the value of the Gilbert damping parameter has to be chosen carefully if the fastest relaxation to the equilibrium spin structure is required. The high effective damping of the breathing mode in the $\alpha\ll 1$ limit (cf. Fig.~\ref{fig4}(a)) ensures that the inverse lifetime of the elliptically polarized excitations remains larger for a wide range of $\alpha$ values in Fig.~\ref{fig6}(b) than what would be expected for circularly polarized modes based on Eq.~(\ref{eqn8b}). Note that contrary to Sec.~\ref{sec3b}, $\textrm{Re}\:\omega_{0,0}$ becoming zero in Fig.~\ref{fig6}(a) does not indicate an instability of the system, since stability is determined by the eigenvalues of the matrix $\boldsymbol{H}_{\textrm{SW}}$ in Eq.~(\ref{eqn2}) which are independent of $\alpha$.

Since the disappearance of $\textrm{Re}\:\omega_{0,0}$ and the bifurcation of $\textrm{Im}\:\omega_{0,0}$ occurs as the excitation frequency becomes smaller, it is expected that such an effect may also be observed at a fixed $\alpha$ value as the external field is decreased. This is illustrated for the $n=0,m=0$ breathing mode of the $2\pi$ skyrmion in Fig.~\ref{fig7} at $\alpha=0.1$. For this intermediate value of the damping, the breathing mode becomes overdamped around $B=0.875\,\textrm{T}$, which is significantly higher than the burst instability between $B=0.775\,\textrm{T}$ and $B=0.800\,\textrm{T}$ (cf. Fig.~\ref{fig1}(b) and the circularly polarized approximation in Fig.~\ref{fig7}(a)). This means that the lowest-lying breathing mode of the $2\pi$ skyrmion cannot be excited below this external field value. In Fig.~\ref{fig7}(b) it can be observed that contrary to the circularly polarized approximation Eq.~(\ref{eqn8b}) following the field dependence of the frequency, for the actual elliptically polarized eigenmode $\left|\textrm{Im}\:\omega_{0,0}\right|$ is almost constant for all field values above the bifurcation point. Although a similar observation was made at the end of Sec.~\ref{sec3c} as the system approached the collapse field at $\alpha=0$, it is to be emphasized again that no instability occurs where $\textrm{Re}\:\omega_{0,0}$ disappears in Fig.~\ref{fig7}(a).

\section{Conclusion\label{sec4}}

In summary, the localized spin wave modes of $k\pi$ skyrmions were investigated in an atomistic spin model, with parameters based on the Pd/Fe/Ir(111) system. It was found that the number of observable modes increases with skyrmion order $k$, firstly because of excitations with higher angular momentum quantum numbers $m$ forming along the larger perimeter of the skyrmion, secondly because of nodes appearing between the multiple domain walls. It was found that the $2\pi$ and $3\pi$ skyrmions undergo a burst instability at low fields, in contrast to the elliptic instability of the $1\pi$ skyrmion. At high field values the innermost ring of the structure collapses in all cases, connected to an instability of a breathing mode.

The effective damping parameters of the excitation modes were determined, and it was found that for the same $n,m$ mode they tend to increase with skyrmion order $k$. The effective damping parameter of the $n=0,m=0$ breathing mode diverges at the burst and collapse instabilities, but no such effect was observed in case of the elliptic instability. For higher values of the Gilbert damping parameter $\alpha$ a deviation from the behavior of circularly polarized modes has been found, with the breathing modes becoming overdamped. It was demonstrated that such an overdamping may be observable in $2\pi$ and $3\pi$ skyrmions for intermediate values of the damping significantly above the burst instability field where the structures themselves disappear from the system.

The results presented here may motivate further experimental and theoretical studies on $k\pi$ skyrmions, offering a wider selection of localized excitations compared to the $1\pi$ skyrmion, thereby opening further possibilities in magnonics applications.

\begin{acknowledgments}

The authors would like to thank A. Siemens for fruitful discussions. Financial support for this work from the Alexander von Humboldt Foundation, from the Deutsche Forschungsgemeinschaft via SFB 668, from the European Union via the Horizon 2020 research and innovation program under Grant Agreement No. 665095 (MAGicSky), and from the National Research, Development and Innovation Office of Hungary under Project No. K115575 is gratefully acknowledged.

\end{acknowledgments}

\end{document}